\documentclass[12pt]{article}
\usepackage[backend=biber,style=authoryear,autocite=inline]{biblatex}

\usepackage{caption}
\usepackage{graphicx}
\usepackage{amssymb} 
\usepackage{amsmath}    %needed for \begin{cases}...\end{cases}
\usepackage{latexsym}   %needed for \Box
\usepackage{bm}
\usepackage{times}
\usepackage{centernot}
\usepackage{float}
\usepackage{color}
\usepackage{wrapfig}
\usepackage{mathtools} %needed for \coloneqq
\usepackage{hyperref}
\usepackage{amsthm}
\usepackage{cleveref}
\usepackage{lineno}
\usepackage{float}

\def\mc{\mathpunct{:}}
\newtheorem{prop}{Proposition}
\newtheorem{lem}{Lemma}
\theoremstyle{definition}
\newtheorem{defn}{Definition}
\begin{document}
\title{The chanciness of time}
  \author{John M. Myers$^{Ia}$ \vspace*{10pt}and  Hadi 
  Madjid$^{IIb}$ \\{\normalsize {\it John A. Paulson School of Engineering and Applied Sciences (Retired),}} \\{\normalsize \it Harvard University, Cambridge, MA 02138$^a$,}\\
  {\normalsize \it 309 Winthrop Terrace, Bedford \vspace*{10pt}MA 01730$^b$.}\\
  {\normalsize jmartmyers@gmail.com$^I$, \vspace*{-.1 in}gailmadjid@comcast.net$^{II}$}}
 \date{}
\maketitle

\begin{abstract}
  Digital network failures stemming from instabilities in measurements of temporal order motivate attention to concurrent events. A century of attempts to resolve the instabilities has never eliminated them.   Do concurrent events occur at indeterminate times, or are they better seen as events to which the very concept of temporal order cannot apply?  Logical dependencies of messages propagating through digital networks can be represented by marked graphs on which tokens are moved in formal token games. However, available mathematical formulations of these token games invoke ``markings''--- global snapshots of the locations of tokens on the graph.  The formulation in terms of global snapshots is misleading, because distributed networks are never still: they exhibit concurrent events inexpressible by global snapshots. We reformulate token games used to represent digital networks so as to  express concurrency.  The trick is to replace global snapshots with ``local snapshots.'' Detached from any central clock, a local snapshot records an action at a node during a play of a token game.   Assemblages of local records define acyclic directed graphs that we call {\it history graphs}.  We show how history graphs represent   plays of token games with concurrent motions,  and, importantly, how history graphs can represent the history of a network operating while undergoing unpredictable changes.
\end{abstract}

\newpage
\begin{center}
{\LARGE The chanciness of time}  
\end{center}

\setcounter{tocdepth}{2}
\tableofcontents
\newpage

\section{Introduction}\label{sec:one}

Digital networks for communications and computation pervade modern life, and scientists visualize networks to explain observable evidence, e.g.\ in sociology, biology, and physics.  In networks, some events depend on others, and so occur temporally one after another, while other events are subject to no determined temporal order; one speaks of {\it concurrent events}.   The mathematical expression that allows for concurrent events is a partial order. A partial order on a set of events orders an event with respect to some subset of events but not necessarily with respect to all the other events. For example, as shown in the figure (b) on the right, A, and B are unordered with respect to C, D, and E. \vspace*{-22pt}
\begin{figure}[H]\hspace*{.2 in}
  %\vspace*{-.3 in}
 \includegraphics[height=1 in]{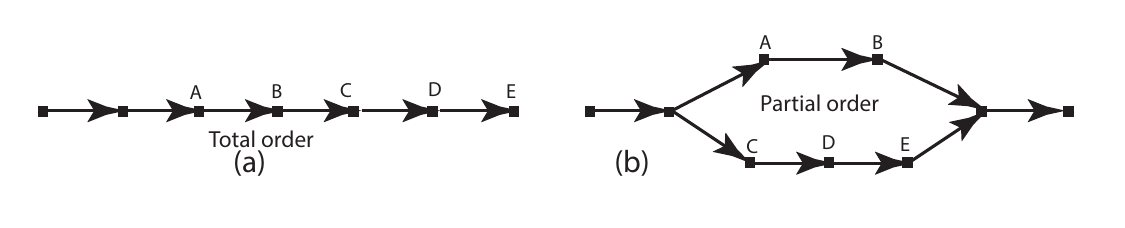}
 \caption*{}
\end{figure}\vspace*{-35pt}
While, in theory, any mathematical partial order  can be extended to a total order \autocite{szpilrajn};  attempts to assign times to concurrent physical events encounter failures. A striking example of trouble in assigning times has never been completely eliminated from digital computers, even after decades of experience.
  A computer takes steps, timed by a clock that ticks (faster than once per nanosecond). 
The computer has a memory, made of tiny flip-flops, into each of which a 0 or 1 can be written, doing away with whatever  bit that was there before.
The computer receives messages from other computers, coded into 0's and 1's.
When a computer receives a 0 or 1 (written on an electronic token), it has to record whether the token came before or after a certain tick of its clock.
So, what happens if the token carrying a 1 arrives just as the clock is ticking and the flip-flop holds a 0?  Does the flip-flop change to 1 or stay with 0?  The simple answer ``It's random.'' turns out to be misleading.  The flip-flop acts as a kind of hinge that could flip one way or flop the other way, so  the hinge can hang up in the middle, teetering on edge, as illustrated below.
\begin{figure}[H]\hspace*{.3 in}
  \includegraphics[height= 1.2 in]{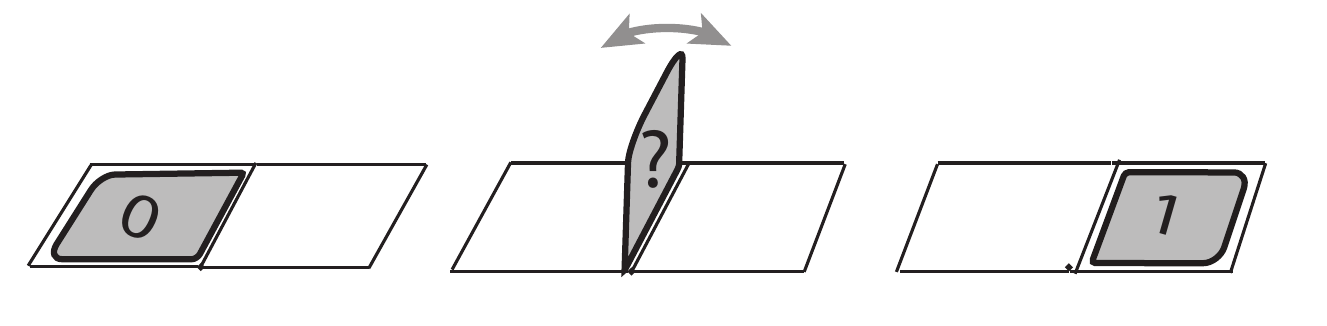}%\vspace*{10pt}
  \caption*{Teetering between "0" and "1".}
  \end{figure}
 In the notorious case of asynchronous  computer interrupts, we and others have measured the way that the instability of this teetering puts different parts of a digital network in conflict, with one part acting on a report that ``$a$ is before $b$'' while another part acts on the report that ``$a$ is after $b$'' \autocite{76arbiter,05aop,19MST}.  The network can then crash.  More recently, we realized that determining temporal order in a three-way race depends on pairwise measurements, leading to a dramatic conclusion.   In close calls  the result for each pair is random.  As a result,  in a three-way race between concurrent events $e_A$, $e_B$, and $e_C$,  a possible result is circular: $e_A < e_B, e_B < e_C$, but $e_C < e_A$, in conflict with the transitivity of temporal ordering or even partial temporal ordering \autocite{19MST}.\footnote{As a principal  advisor to the National Institute of Science and Technology has put it, ``time'' as used in science and technology is an artifact \autocite{allan87,14aop}. 
}

Since the early days of distributed computing, two attitudes toward concurrent events in networks continue to contend.  Abstracting out of sight the instability of decisions of temporal order, some authors  assume that each event must happen ``at some time.''
Although this artificial ``time''  underpins valuable techniques in computer science \autocite{19Lamport}, we seek an alternative way to represent events of a digital network that recognizes that the concept of a total temporal order is inapplicable to concurrent events.

To model concurrency in networks, we modify the mathematical definitions of
token games played on marked graphs.  Mathematical formulas used to represent
motion do not themselves move; they sit still
on the page; how something still can represent something moving seems mysterious until it strikes us that formulas represent {\it snapshots} of that
motion.  The current mathematics of marked graphs invokes
``snapshots,''  termed {\it markings}.  A marking expresses
the location of tokens at a moment of a play of a token game.  If the players moved one at a time, as in chess, a play of the game could be recorded as a sequence of
snapshots taken between moves, in moments when the positions of pieces on a game
board are still.  A marking of a marked graph presupposes such a snapshot, but
for token games expressing a digital network, the moves are not ``one at a time,'' and there need be no moment for a snapshot to show a  \vspace*{8pt} marking.\\
\setlength{\fboxrule}{1.2pt}
\fbox{\begin{minipage}{36em} {\it \bf 
The specific problem we tackle here is to formulate a mathematics of token games that exhibit concurrency without invoking markings.}
\end{minipage}} \vspace*{10pt}\\

We get a hint about how to proceed from an intuition of something missing in mathematical physics. In 2005 we proved that quantum theory exposes a gap between evidence and any explanations of that evidence, a gap bridged only by reaching beyond logic to make a guess \autocite{05aop}, implying  an  unpredictability more drastic than the uncertainty principle:
Any explanation of evidence consistent with quantum mechanics is  vulnerable to a surprise from future evidence; in brief ``no final answers.''   And, the need to guess gives license for the imagination in science.  With this hint, we seek mathematics not to predict but to express real or imagined records of what has been measured, uncluttered by any attempt to predict future behavior.  The focus on real or imagined records leads to a second step that provides an alternative to markings as global snapshots: We formulate the mathematics of networks in terms of {\it local records} of individual events of a network, made without reference to any global clock.  I.e., rather than using the global snapshots that are markings, we build on local snapshots.

We represent assemblages of local records of plays of token games by use of graphs that
we call {\it history graphs}.\footnote{In the context of computer science, these are directed graphs, variously called `event histories' \autocite{19Lamport}, `occurrence graphs' \autocite{holtOcc} `occurrence nets' \autocite{79PetriB}, \autocite[p.\ 61]{88BestFer}, and `histories' \autocite{82Petri}.}
Thus, two distinct types of graph enter the
 story: (1) game graphs (reminiscent of a chess board) on which token games
 are played, and (2) history graphs expressing records of plays of token games.
 To make the distinction clear, we speak of a game graph as consisting of nodes
 connected by arrows, and of a history graphs as consisting of events connected
 by edges.  An arrow in a game graph illustrates a channel for a token to
 propagate from one node to another.  In contrast, an edge in a history graph represents the local record of the arrival of a token.

In its basic form, a history graph is an acyclic  directed graph, consisting of edges (directed) and vertices that we speak of as `events'.
As applied to token games, the events and edges of a history graph express dependencies of recorded moves of the game on prior moves.  
There are two advantages of history graphs over markings as a basis for the mathematics of token games.  (1) They enable the definition of concurrent events as those unconnected by a path  in either direction---a way to say that concurrent events are not only unordered, but are  unorderable. (2) They can express records of the operation of a network that continues while its configuration undergoes predictable or unpredictable changes.  

In \cref{sec:2} we introduce history graphs along with their specializations appropriate to representing plays of token games.
In \cref{sec:3} we discuss token games played on game graphs to represent
computational networks and their unpredictable changes, and we show a token
game in which there can be no global snapshots.  In \cref{sec:4} we show how
history graphs of plays of token games allow the study of these games to be based on local snapshots, without
resort to global snapshots.  In basing the mathematics of token
games on local records, we highlight the acts of visualization
necessary to interpret local snapshots as representing motion, acts reminiscent of
those needed to infer the behavior of animals from tracks left in the sand.  

\section{Local records of interdependent events}\label{sec:2}
\begin{figure}[H]\hspace*{.4 in}
  %\vspace*{-.3 in}
 \includegraphics[height=4.6 in]{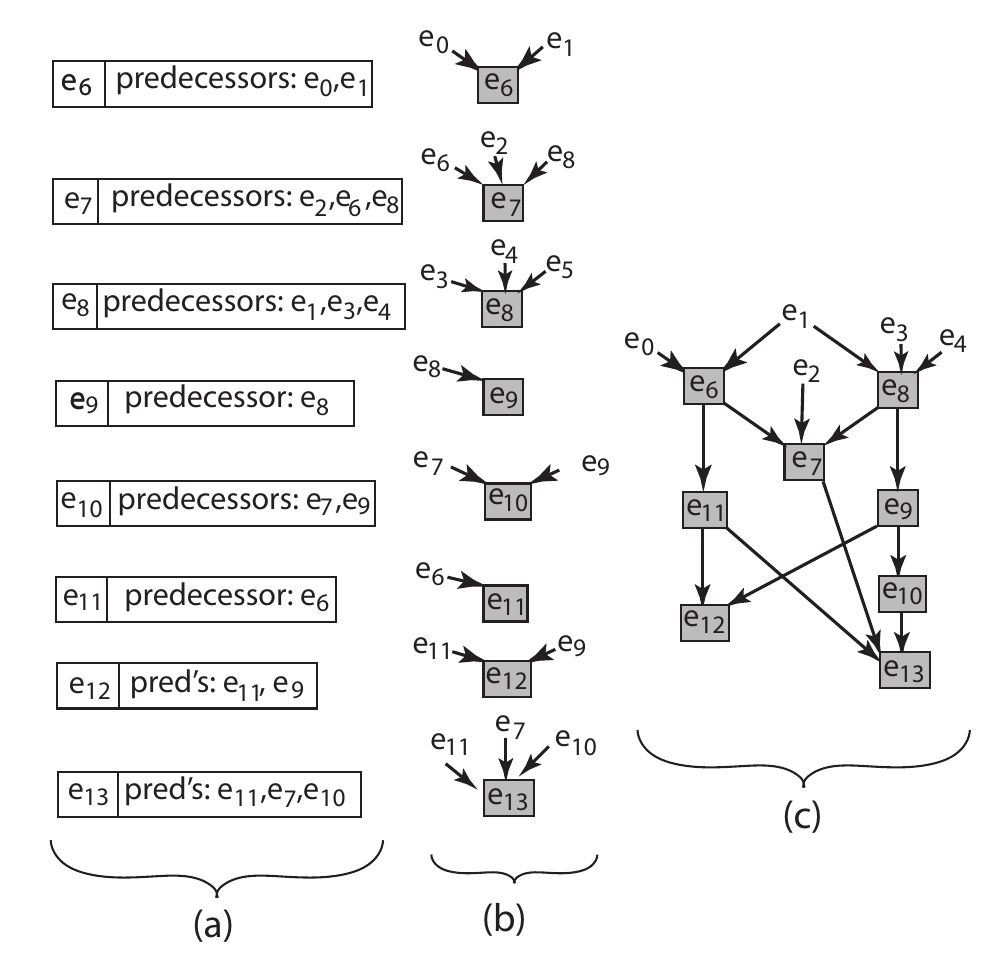}
 \caption{(a) Local records of event predecessors; (b) graphical portrayal of local records; (c) history graph assembled from local records.\label{fig:pic} }
\end{figure}

We imagine local records from which to assemble the dependencies among events.
A record of an event includes information that comes to it from other events.  An event depends directly on other events that we call its {\it predecessor} events: If event $e$ is a predecessor of event $e'$, we draw a directed edge from $e$ to $e'$ ($e \rightarrow e'$); also we speak of $e'$ as a {\it successor} of $e$. Suppose that  the record for each event names that event and also its predecessor events, as in \cref{fig:pic}(a).  Each such record has a graphical representation, as in \cref{fig:pic}(b)
By tracing back predecessors of predecessors, the local records define a larger history graph, illustrated in \cref{fig:pic}(c).  

 A history graph is acyclic, and hence defines a partial order on its events.  A history graph  allows us to define {\it concurrency} as a relation between pairs of events: $e$ and $e'$ are concurrent events if there is neither a directed path in the history graph from $e$ to $e'$ nor a directed path from $e'$ to $e$.  Long ago, Petri gave the same definition in the context of Petri nets, and did this with the same motivation of defining concurrency without reference to clock times \autocite{79PetriB}.\footnote{In contrast, some other authors speak of concurrent events as happening at undetermined times,
 which to us vitiates the main conceptual advance of Petri nets as expressing relations without resort to the concept of time.} In contrast to events as points in a spacetime manifold, a history graph indicates nothing  about spatio-temporal extent; one event might be fast, another slow, one confined, another spatially extensive.  

History graphs involve choosing a level of granularity of description.  For example, a coarsening of history graph in \cref{fig:pic}(c) can combine events $e_6$, $e_7$, and $e_8$ into the single event $e_{\displaystyle *}$.  A coarsening of a history graph is expressed by a
 mapping  of events and edges  that breaks no edges and creates no cycles.
 \begin{figure}[H]\hspace*{.2 in}
  %\vspace*{-.3 in}
 \includegraphics[height=2.8 in]{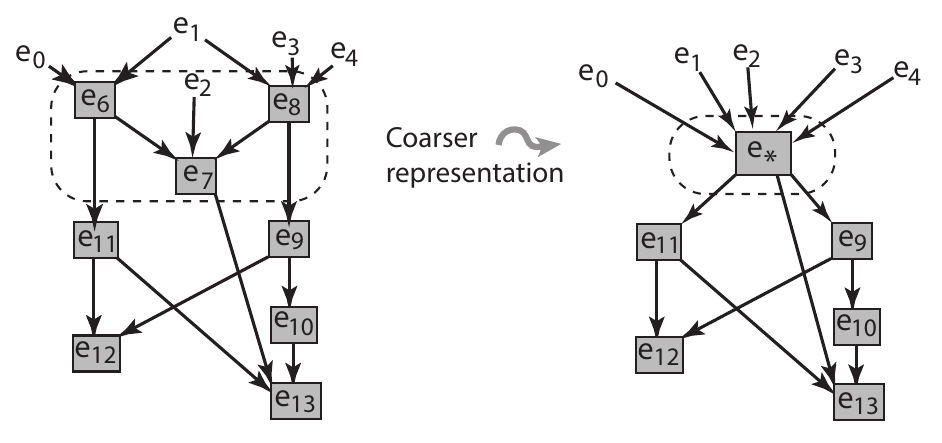}
 \caption{Coarsening that combines $e_6$, $e_7$, and $e_8$ into the single event $e_{\displaystyle *}$\label{fig:coarse} }
\end{figure}

 Later on, we will consider records that include not just the names of predecessors, but also what they supply.  For example, a predecessor of an event of calculation can supply a label consisting of a string of bits  (0's and 1's).  Multiple history graphs can pertain to a given situation; for example, a smaller earlier history graph might be a subgraph of a larger history graph that expresses additional local records.

\subsection{History graphs for networks of nodes}

\begin{figure}[H]\hspace*{1.5 in}
  %\vspace*{-.3 in}
 \includegraphics[height=4.5 in]{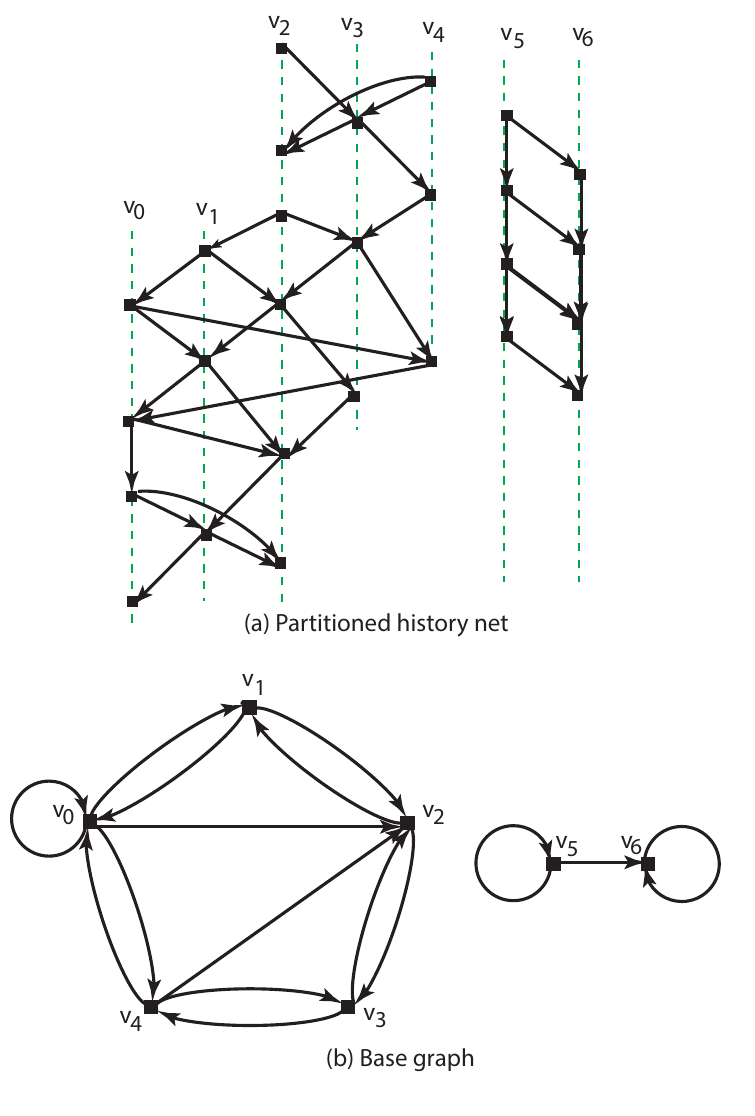}
 \caption{(a) history graph with events partitioned into columns for nodes;
   (b) base graph (explained below).\label{fig:examp} }
\end{figure}

History graphs can represent locally recorded communication activity in a
network. Then each event of the history graph represents one of the sequence of
actions of a node of the network, so that the events of the history graph are
partitioned into sets, one set of each node of the network. We speak of {\it partitioned history graphs}. We assume a
finite set of nodes of the network.  In our pictures of history graphs, we show events, drawn as small black squares,  arranged in columns, indicated by vertical dashed lines in green, one column for each node of the network.
\Cref{fig:examp} shows an example. 

For an event in a partitioned history graph, we speak not only of predecessor events and successor events, but less specifically of predecessor nodes and successor nodes:
\begin{defn}
  A predecessor node of an event in a partitioned history graph is the column containing the corresponding predecessor event, and the successor node of an event in a partitioned history is the column containing the corresponding successor event.
\end{defn}
In a partitioned history graph for a network undergoing changes during its  operation, two events in the same column can differ in their predecessor and successor nodes.

\subsection{Base graph of a partitioned history graph}\label{sec:2.2}
For any partitioned history graph there is an associated {\it base graph}, defined as follows. The base graph has a node for each column of events of the history graph and an arrow between nodes if and only if the partitioned history graph has a corresponding edge.

As discussed in \cref{sec:A} of the Appendix, we say an arrow of a game graph connects a source node to a target node, and an edge of a history graph connects a source event to a target event.
We call a mapping from a subgraph of a partitioned history graph to its base a  {\it projection} if and only if:
\begin{enumerate}
\item  Each edge in  the domain maps to the  arrow of the base that has the source node and target node of the edge.
\item Each event in the domain maps to the node of the base for the column of that event;
\end{enumerate}
Although the partitioned history graph is acyclic, in most cases the base
graph has cycles; it can also have loops, as shown in \cref{fig:examp}(b).
(An edge connecting an event of a column to the same column in a partitioned history graph appears if and only if there is a loop connecting the corresponding node of the game graph to itself.)

Connectivity of base graphs is of interest.  For example, consider \cref{fig:examp}.
Because no edges bridge the columns for $v_5$ or $v_6$ to the other node columns of the partitioned history graph, the base graph in \cref{fig:examp}(b) is disconnected. The component for $v_0$, $v_1$, $v_2$,$v_3$, and $v_4$  is strongly connected, while the component for $v_5$ and $v_6$ is weakly but not strongly connected: there is no directed path from $v_6$ to $v_5$.  That distinction between strongly and weakly connected components of a base graph will matter when we come to live and safe token games in \cref{sec:4.1}.  

In the next section we introduce token games on marked graphs---mathematically defined games that, for example, can simulate digital computations done in a distributed network.  After that, in \cref{sec:4}, we show how local records assembled into history graphs allow the mathematical formulation of token games representing networks, without assuming any global state of a network---i.e.\ without markings.

\section{Token games}\label{sec:3}
\begin{figure}[H]\hspace*{1.8 in}
  %\vspace*{-.3 in}
 \includegraphics[height=1.9 in]{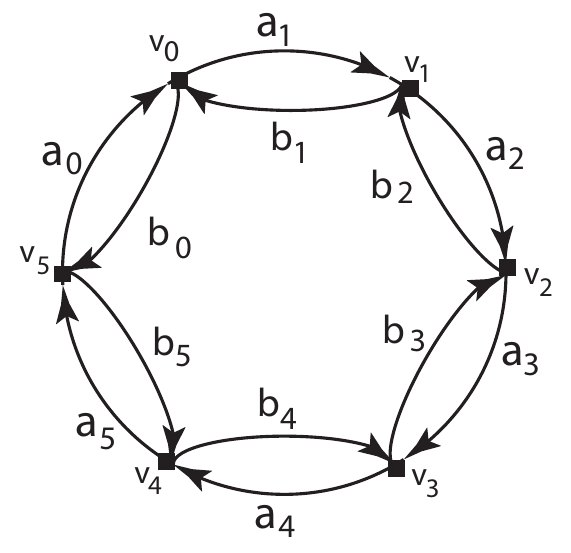}
 \caption{The ``Hex'' game graph.\label{fig:game} }
\end{figure}

A token game straddles mathematics and physical behavior.  It is played by one or more players on a game board, according to rules, and yet the rules are sufficiently clear to be represented mathematically.  The game board implements a visual image of a game graph, with a node for each player and arrows that express communication channels from node to node.\footnote{As discussed in \cref{sec:A},
the game graphs can have multiple arrows and loops.} As illustrated in \cref{fig:game}, unlike a history graph, a game graph can contain \vspace*{-10pt}
cycles.

A token game calls for a player for each node of a game graph. Each player for a node removes tokens and places tokens on arrows incident with the node, according to rules.
 First is the \vspace*{8pt}
following\\
\setlength{\fboxrule}{1.2pt}
\fbox{\begin{minipage}{36em} {\it \bf Firing rule: On occasion a token will reside on each of the arrows pointing into a node; then the player removes those tokens and places a token on each of the arrows that point away from the node.}
\end{minipage}} \vspace*{10pt}\\
That is the only rule for token games on marked graphs as defined in \autocite{71MG}. However, we want to expand token games to express computations that take place in a digital network, and for this we need to allow for labels on tokens, along with labeling rules for generating output labels from input labels. 
For token games with labeled tokens,  there is a {\it labeling rule} for each node that specifies labels on outgoing tokens as functions of labels on incoming tokens; this  labeling rule usually differs from node to node.
  
In a token game that computes, a node of the game graph represents a device, e.g.\ a logic gate, that can repeatedly execute a logical operation.  The firing of a node with its labeling rule then simulates an occurrence of the logical operation.\footnote{Token games representing computations can be played on Petri nets without the need to label tokens.  While the simplicity of unmarked tokens is appealing, there is a cost in that the graphs are more complex, and the token games on Petri nets involve choices  of where a token goes. Marked graphs are a simplification of Petri nets in which there are no such choices.  Labeling the tokens regains the variations needed to model computations, while preserving the simpler graphs.  \cite{82Johnsonbaugh,SRL}
} However, when we are focusing on the dependencies of one firing on another firing,  we often ignore the token labels. In our own work, we play  small token games to gain insight, while the mathematics supports propositions that apply to token games on any finite game graph.

For a token game to represent the logic of computations, two restrictions are necessary; the game must be {\it live} and {\it safe}.
\begin{defn}
We call a token game {\it live} if all of its nodes can be fired repeatedly, without end.  
\end{defn}
 
\begin{defn}\label{defn:safeDef}
We call a token game {\it safe} if no more than one token can ever be placed on any arrow.  
\end{defn}
In the following, we concentrate mostly on token games that are live and safe.

\subsection{Token game without any still moment for a global snapshot}\label{sec:no-snap}

\begin{figure}[H]\hspace*{1.2 in}
  %\vspace*{-.3 in}
 \includegraphics[height=3.5 in]{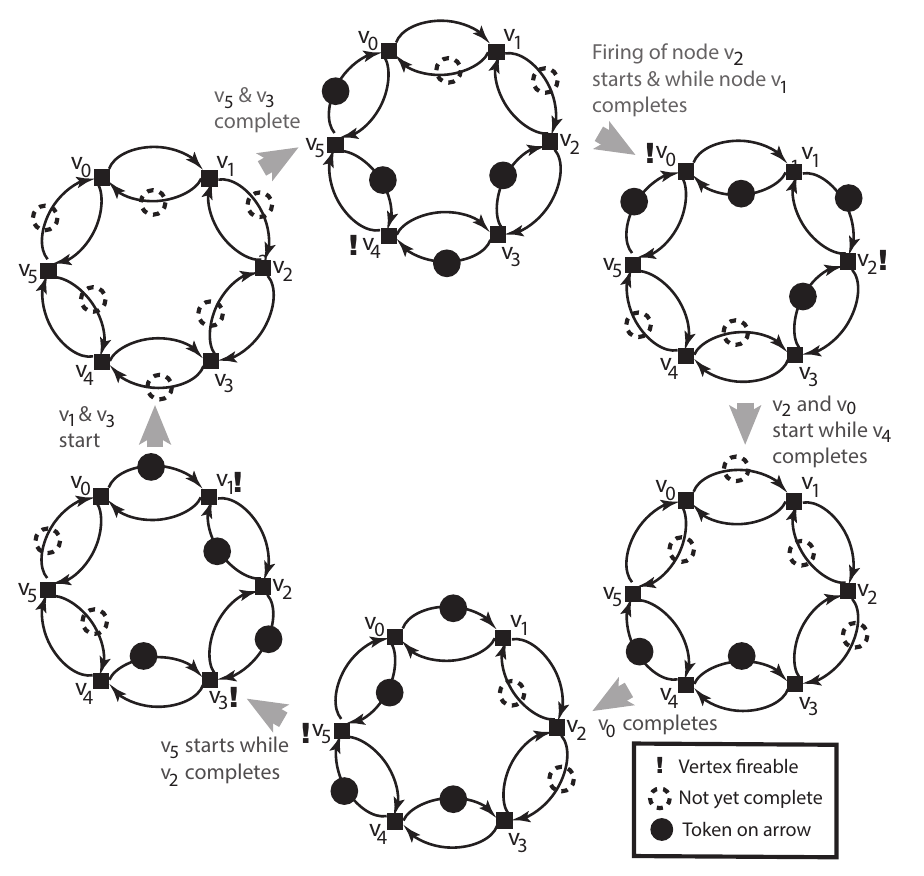}
 \caption{Play of a ``Hex(3,3)'' game with no still moment.\label{fig:no-still} }
\end{figure}
\Cref{fig:no-still}  illustrates a play of a token game with unlabeled tokens, satisfying the firing rule, in which each node has still moments for local snapshots, but there is no ``still moment'' for all the nodes at once for a global snapshot to  show unambiguous positions of tokens.  The six pictures represent global photographs attempting to catch markings.  The small dotted circles are added to the photographs to show places on arrows where no tokens reside but which will eventually receive tokens from a firing that is in progress  when the picture is taken.  However, when those arrows receive tokens, other tokens will have been removed from other firings that have begun but are not yet complete. The six pictures can be followed clockwise around and around, illustrating continuous play with no moment when all tokens are at rest.  It is the absence of any global ``still moment'' that motivates us to define marked graphs without invoking the notion of  markings as global snapshots.

\subsection{Local records of three token games on the Hex game graph}
\Cref{table:1} shows local records that define partitioned history graphs for three distinct token games that can be played on the ``Hex'' game graph of \cref{fig:game}. 
\begin{table}[H]
  \centering
\begin{tabular}{||l|l|l||}\hline
 \;\;\;Hex(1,5)&\;\;\;Hex(2,4)&\;\;\;Hex(3,3)\\\hline
  $v_0\mc 1|v_5\mc 0;v_1\mc 0$ &  $v_0\mc 1|v_5\mc 0;v_1\mc 0$&$v_0\mc 1|v_5\mc 0;v_1\mc 0$\\
  $v_1\mc 1|v_0\mc 1;v_2\mc 0$&$v_1\mc 1|v_0\mc 1;v_2\mc 0$&$v_1\mc 1|v_0\mc 1;v_2\mc 1$\\
  $v_2\mc 1|v_1\mc 1;v_3\mc 0$&$v_2\mc 1|v_1\mc 1;v_3\mc 1$&$v_2\mc 1|v_1\mc 0;v_3\mc 0$\\
 $v_3\mc 1|v_2\mc 1;v_4\mc 0$&$v_3\mc 1|v_2\mc 0;v_4\mc 0$&$v_3\mc 1|v_2\mc 1;v_4\mc 1$\\
$v_4\mc1|v_3\mc1;v_5\mc0$&$v_4\mc1|v_3\mc1;v_5\mc0$&$v_4\mc1|v_3\mc0;v_5\mc0$\\
$v_5\mc1|v_4\mc1;v_0\mc1$&$v_5\mc1|v_4\mc1;v_0\mc1$&$v_5\mc1|v_4\mc1;v_0\mc1$\\
$v_0\mc2|v_5\mc1;v_1\mc1$&$v_0\mc2|v_5\mc1;v_1\mc1$&$v_0\mc2|v_5\mc1;v_1\mc1$\\\hline
 \end{tabular}
\caption{Local records for three token games on Hex game graph.}
\label{table:1}
\end{table}

A node can fire more than once, so the event name includes not only the node that fires but also an index that is incremented each time the node fires.  The form of a local record is then
[node name, index$|$ node name, index; node name, index;\ldots ],
where events to the right of the ``$|$'' are predecessor events.  For example, \cref{table:1}
shows local records for  plays of three distinct token games that can be played on the game graph shown in \cref{fig:game}.  The nodes of the graph are $v_0, v_1,v_2,v_3,v_4$ and $v_5$.  The first record of the play of the game denoted by ``Hex(1,5) is [$v_0\mc 1|v_5\mc0;v1\mc0$] .  That record gives the predecessors of the event $v_0\mc 1$ as $v_5\mc 0$ and $v1\mc0$.  The numbers after the colons are indices.  The order in the list of records for each token game is arbitrary; in general the node firings are  partially ordered, as determined by the indices. Note that the indices of any node indicate only the total order of firings of that node, so that any integer added to all the indices for a node has no significance.

\section{Replacing markings by local records and partitioned history graphs}\label{sec:4}
Now we turn to the main business of formulating the mathematics to represent
digital networks without invoking global snapshots.  For this representation, we
formulate the mathematics of token games in terms of the partitioned history
graphs that express local records of plays of these games.  In building on
history graphs, one needs to replace markings by local records of plays of token
games.  Second, one needs to retool ones imagination by which to bridge the gap
between mathematics and its application to descriptions of motion: away from the
tacit assumption of global snapshots to learning to think in terms of partitioned history graphs.

Partitioned history graphs express the local records of node firings in plays of a token
game, e.g. the local records shown in \cref{table:1}. Each firing event in a
play of a token game is expressed in the history graph by an event with its
in-pointing arrows.  As formulated in \cref{sec:2.2}, the partitioned history graph for a play of a
token game has the game graph as its base graph:  the partitioned history graph
projects to the game graph, so the events of each column project to
to the corresponding node of the game graph. Successive events on a given column
of a partitioned history graph correspond to successively generated local records of firings
of the node represented by the column.  Those records are indexed: 0, 1,
2, \ldots Although it constrains the history graph, in most cases the game graph
leaves open the possibility of more than one history graph, corresponding to
more than one token game that can be played on a given game graph.
 We focus first on history graphs for live and safe token games; later, in \cref{sec:5},
 we will examine history graphs that can express unpredictable changes of the game graph that occurs during continuing play.

 \subsection{History graphs for plays of live and safe token games}\label{sec:4.1}
 To begin,  the local records of a play of a live and safe token satisfy  the following 
\begin{prop}\label{prop:incr}    
From one firing to the next firing of any node in a live and safe token game, the indices on local records for that node and  its  predecessor nodes all increase by 1. 
\end{prop}
\begin{proof}

For any node $v$ of  the game graph with node $w$ as a predecessor node, consider local records of any two node-firing events $v\mc m$ and $v\mc n$.   Then
the local records of firing events for $v$ show [$v\mc m| w\mc m';$\ldots] and
[$v\mc n| w\mc n';$\ldots]. We argue that $n-m=n'-m'$.  If not,  there must be a different number of firings of $v$ than of $w$ connecting the two records of $v$ firings.  But such a difference would require  a $w$-firing that was not a predecessor for a $v$-firing or a $v$-firing without a $w$-firing as a predecessor event, both of which are precluded  in a live and safe token game by the firing rule. Because consecutive firings of a node increment its firing index by 1, the indices of its predecessor-node firings are also incremented by 1.
\end{proof}
An example of this increase of indices can be seen by comparing the bottom record in each column of \cref{table:1} to the top record.

\begin{enumerate}
\item From \Cref{prop:incr}
it follows that the partitioned history graph of a live and safe token game is repetitive: it can be constructed by pasting together a sequence of copies of a fragment that contains one event from each column. One of the many ways of constructing such fragments is shown in \cref{fig:3game}. We tacitly assume that the history graphs for live and safe token games have many such repetitions.
\item  In contrast to the changes discussed in \cref{sec:5}, for a token game played on a fixed game graph, the game graph is the base graph of the partitioned history net.  I.e., the partitioned history graph projects onto  the game graph.  The projection of the history graph to the game graph takes every event in a column on the history graph to the corresponding node in the game graph, and takes every edge of the history graph to the corresponding arrow of the game graph.
\item Important to explaining the relation between partitioned history graphs and plays of token games are paths in history graphs.   If a path includes a firing event of a game-graph node $v$, we say the path {\it
  touches} the node $v$.
\item Of special interest to the study of live and safe token games are paths in history graphs
  that project to cycles in the game graph.  We call such paths {\it spirals}.
  For each cycle in the game graph, the  history graph contains a spiral that projects to that cycle.   The spiral  repeatedly touches each node of the cycle, following the cyclic order of the cycle.
\end{enumerate}
    \begin{defn}
The {\it pitch} of a spiral is the increase in the index of successive events on a column that it touches.
\end{defn}
    Pitches of spirals are illustrated in \cref{sec:4.4}
    
Item one above can be strengthened to:
\begin{prop}\label{prop:sector}
The events of a partitioned history graph of a continuing play of a live and safe token game can be partitioned ``crosswise'' to the columns into a sequence of subsets that we call {\it transverse sectors}, such that:
\begin{enumerate}
\item Each transverse sector contains 1 event of each column, and
  \item There is no path in connecting two events in a transverse sector that includes any event not in the sector.
\end{enumerate}  
\end{prop}
(\Cref{prop:sector} expresses in local physics the content that in expressed by markings in {\scshape Theorem} 7
of \autocite{71MG}: ``Let M be a live
marking. There exists a firing sequence leading from M to itself, in which every
vertex fires exactly once.'')

 \subsubsection{Example}
\begin{figure}[H]\hspace*{.8 in}
  %\vspace*{-.3 in}
 \includegraphics[height=3.5 in]{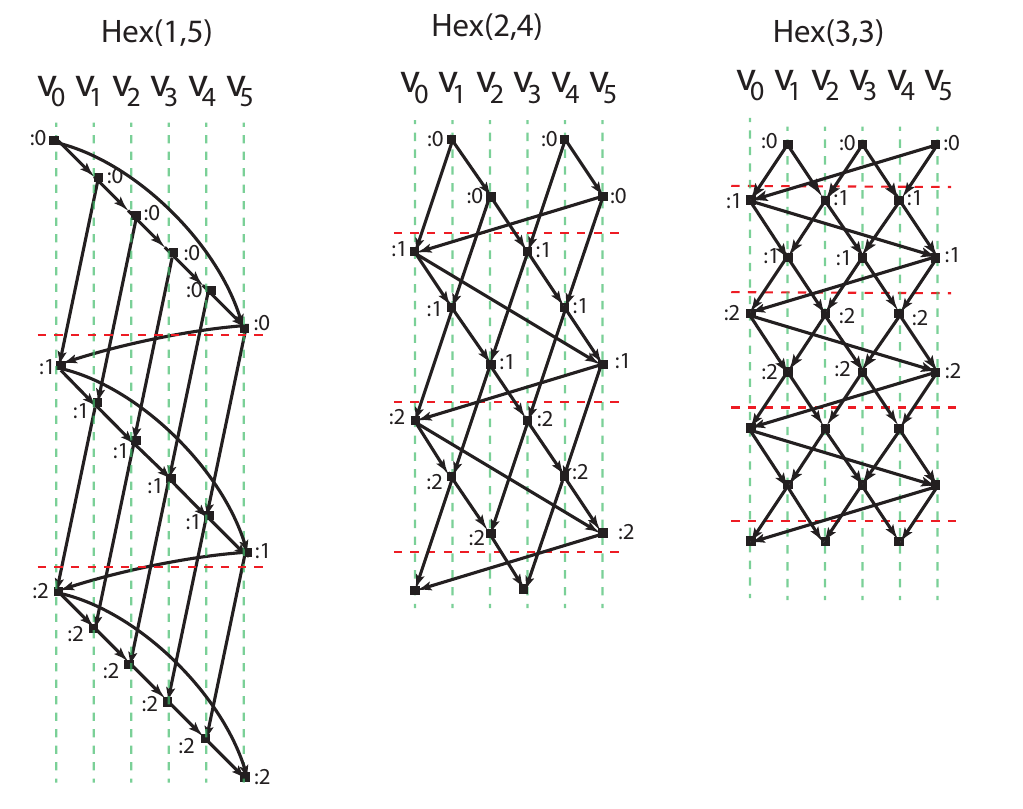}
 \caption{History graphs for three live and safe token games on the ``Hex'' game graph.\label{fig:3game} }
\end{figure}

The history graphs for the three token games of \cref{table:1} are shown in \cref{fig:3game}.  Each dashed vertical line defines a column  of local records of firing events of one node of the game graph. Along the column for a node of the game graph, the indices of the records advance by 1 for each successive event of the node.  The indices on the events correspond to those in \cref{table:1}.
Events in the history graphs shown in \cref{fig:3game} have names of the form $v_j\mc k$ where $j \in \{0, 1, 2, 3, 4, 5\}$ indicates which node of the ``Hex'' game graph and $k$ is an index that is incremented for successive firings of a node.  
The red dashed lines show possible partitions of events into transverse sectors.

\begin{figure}[H]\label{fig:unsafeFig}
  \includegraphics[height=3.5 in]{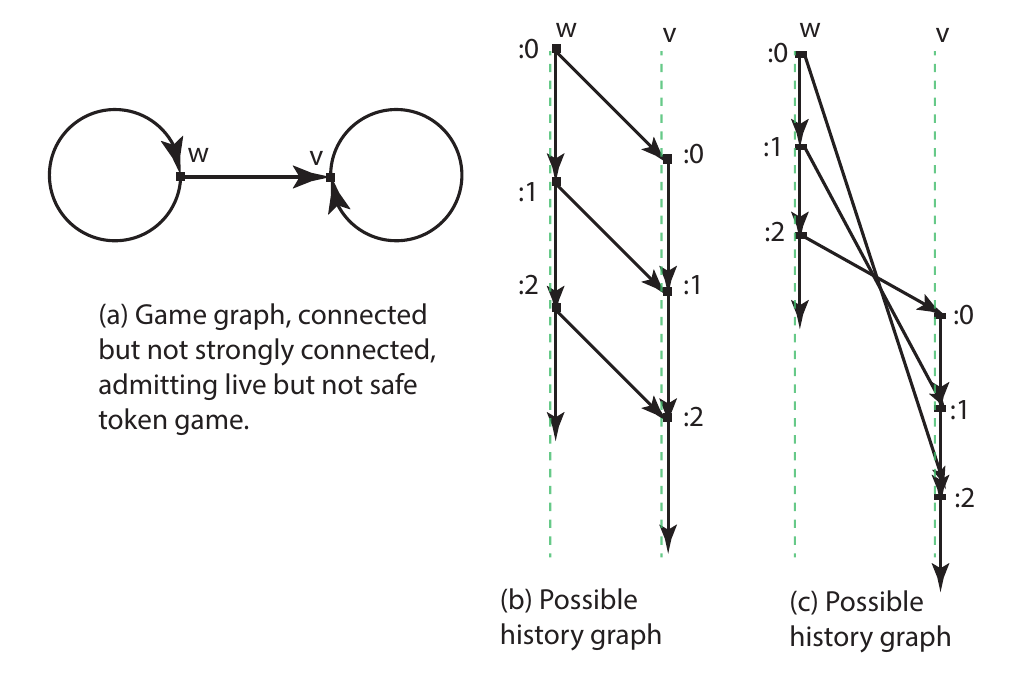}
  \caption{Disordered events in a history graph for play of a token game that is live but not safe.}
  \end{figure}

The hypothesis that the token game be not only live but also safe is essential to \Cref{prop:incr}. As illustrated in \cref{fig:unsafeFig}, history graphs (b) and (c) are both possible for a play of a token game on the game graph (a), and history graph (c) violates the ordering that holds for history graphs of token games that are not only live but also safe.

\begin{prop}\label{prop:safeHist}
  If a live token game is safe, and its  history graph shows two successive firing events of a node $v$ of the game graph,  there are directed paths from the first $v$-firing event  through a firing event of each of its successor events and on to the second $v$-firing event.
\end{prop}
\begin{proof}
  For a token game to be safe, the node $v$, having fired, cannot fire a second time before tokens have been removed from all its arrows for which it is the source node.  This implies that the successor nodes of $v$ must fire, removing tokens from the outgoing arrows of $v$, before that $v$ can fire again.  Thus if  $e_v$ is a firing event of $v$ and $e'_v$ is the next firing event of $v$, each successor event of $e_v$  is contained is in a path from $e_v$ to $e_v'$.
\end{proof}

Nest we come to a purely graphical property pertinent to game graphs.
\begin{lem}\label{lem:arrow}
  If every arrow of a finite connected graph is contained in a directed circuit, the graph is strongly connected.
\end{lem}
\begin{proof}
    For any two vertices $v$ and $w$, for some $n$ there is a weak path (arrow direction ignored) from $v$ to $w$ consisting of arrows $a_i,\; i= 1,\ldots, n$.  Construct a directed path from $v$ to $w$ by concatenating $n$ directed paths as follows.  For $i \in \{1,\ldots,n\}$, if $a_i$ points in the direction from $v$ to $w$, then path $P_i$ is $a_i$ along with its incident nodes.  If $a_i$ points in the opposite direction, $P_i$ is the path obtained from any cycle to which $a_i$ belongs by deleting $a_i$.  The concatenation of the $P_i$ is then a directed path from $v$ to $w$.
\end{proof}

With this lemma, we prove a substitute for {\scshape Theorem} 3 of \autocite{71MG}  without resort to markings. 
\begin{prop}\label{prop:strongG}
    For a token game to be live and safe, it is necessary that each connected component of the game graph be strongly connected.
\end{prop}
\begin{proof}
  If a component of the game graph is not strongly connected, then by \cref{lem:arrow}, it contains an arrow not in any circuit, so the component contains no path from the target of the arrow to its source.  If the graph admitted a live and safe token game, then in that game the source node of that arrow would be repeatedly fireable.  Then, for lack of a return path, it could fire repeatedly without requiring any firing of the target node, making the token game unsafe.
\end{proof}

Along the way, we have shown that every arrow of any graph that admits a live and safe token game belongs to a circuit, which is
Corollary 2 of \autocite{71MG}, again without resort to markings.

\Cref{prop:strongG} has an implication for the use of marked graphs to represent \vspace*{8pt}computations.
\\
\setlength{\fboxrule}{1.2pt}
\fbox{\begin{minipage}{36em} {\bf To represent a computation by a live and safe token game, one must have a strong game graph; a strong game graph allows no source (a node with not arrow pointing into it) and no sink (a node that lacks an arrow pointing out from it).  If a computation is presented as having a start and an end, suggesting source(s) and sink(s), we reframe the computation as in process-control: i.e.,  we augment the  game graph with an ``ENV'' node for an environment, along with an arrow from each end node to this ENV node, and with an arrow from the ENV node to each start node \autocite{SRL}.}
\end{minipage}} \vspace*{6pt}\\

The translation from live and safe token games to history graphs can now be summarized:
\begin{prop}\label{prop:LS}
A partitioned history graph represents a play of a live and safe token game if and only if the history graph has the following properties:
1. There is a projection of the partitioned history graph to a strongly connected game graph;
2. For each column of the partitioned history graph, there is a path that contains all the events of that column, and
3. Each edge of the partitioned history graph belongs to a spiral of unit pitch.  
\end{prop}

Next, we turn to a known result based on markings as global \vspace*{6pt}snapshots.\\ 
{\scshape Theorem} 4 of \autocite{71MG}
 ``For every finite, directed, strongly connected graph there exists a live and safe
\vspace*{6pt}marking.''\\
Rephrased without use of markings, the equivalent content can be expressed by:
\begin{prop}\label{prop:LShist}
  Every finite, directed, strongly connected graph is the base graph of a partitioned history graph such that:
1. For each column of the partitioned history graph, there is a path that contains all the events of that column, and
2. Each edge of the history graph belongs to a spiral of unit pitch.  
\end{prop}
From \cref{prop:LS} we claim that \cref{prop:LShist} can be proved, and we also claim it can be proved purely in graph theory, with no reference to markings or even to token games.  So far, however, we have not proved it in this way, and invite our readers to do so.

\Cref{prop:LShist} implies:
\begin{prop}
  On every finite, directed, strongly connected graph a live and safe token game can be played.
\end{prop}

\subsection{Game graphs that support more than one live and safe token game}
Using the now disparaged markings, the properties of a game graph necessary and sufficient for its admitting more than one live and safe token game have been established \autocite{80Murata,SRL}.  
\begin{defn} 
If a  game graph contains $m\ge 3$ nodes that all belong to two circuits that order the nodes in opposite cyclic orders, we say the game graph has {\it opposing circuits} through these nodes.
\end{defn}
Translated into local records and history graphs, propositions proven in \autocite{80Murata,SRL}
can be stated as:
\begin{prop}\label{prop:murata}
If a strongly connected directed graph contains opposing directed circuits it
admits more than one live and safe token game.
\end{prop}
The direct proof of \Cref{prop:murata} without resort to markings is left to the future.

\subsection{Index increments and logical distance in token games}
In physics, Einstein defined distance between two local clocks $A$ and $B$ by the local time required for a clock $A$ to transmit a signal to $B$ and to receive a return signal from $B$.  This definition requires a metric, namely the time as told by clock $A$.  A topological analog comes naturally to marked graphs.  For a live and safe token game, the {\it logical distance} between a node $v$ of a game graph and a node $w$ is the number of firing events of $v$ for a label put on a token and passed on through nodes of the shortest path to $w$ and back again on the shortest path for $w$ to $v$
\autocite{SRL}.  The logical distance depends not only on game graph but also on the token game, which in earlier work we expressed by markings, now to be avoided.

Without resort to markings, the logical distance can be read from the partitioned history graph of a play of the token game as follows.   Two concatenated paths are involved.
\begin{enumerate}
\item $P_1$ is the shortest path in the partitioned history graph from an event $e_v$ in the column of events for $v$ to an event $e_w$ in the column for $w$; 
\item $P_2$ is the shortest path from the $e_w$-event back to the column of $v$ events, arriving at an event $e'_v$.
\end{enumerate}
The logical distance $D(v,w)$ is the increment in the indices of $v$-events between $e_v$ and $e'_v$.  On the history graph this is 1 more than the number of $v$ events between $e_v$ and $e'_v$.  In \autocite{SRL}, logical distance is shown to satisfy the axioms of distance---triangle inequality etc.

\begin{figure}[H]%\hspace*{.8 in}
  %\vspace*{-.3 in}
 \includegraphics[height=4 in]{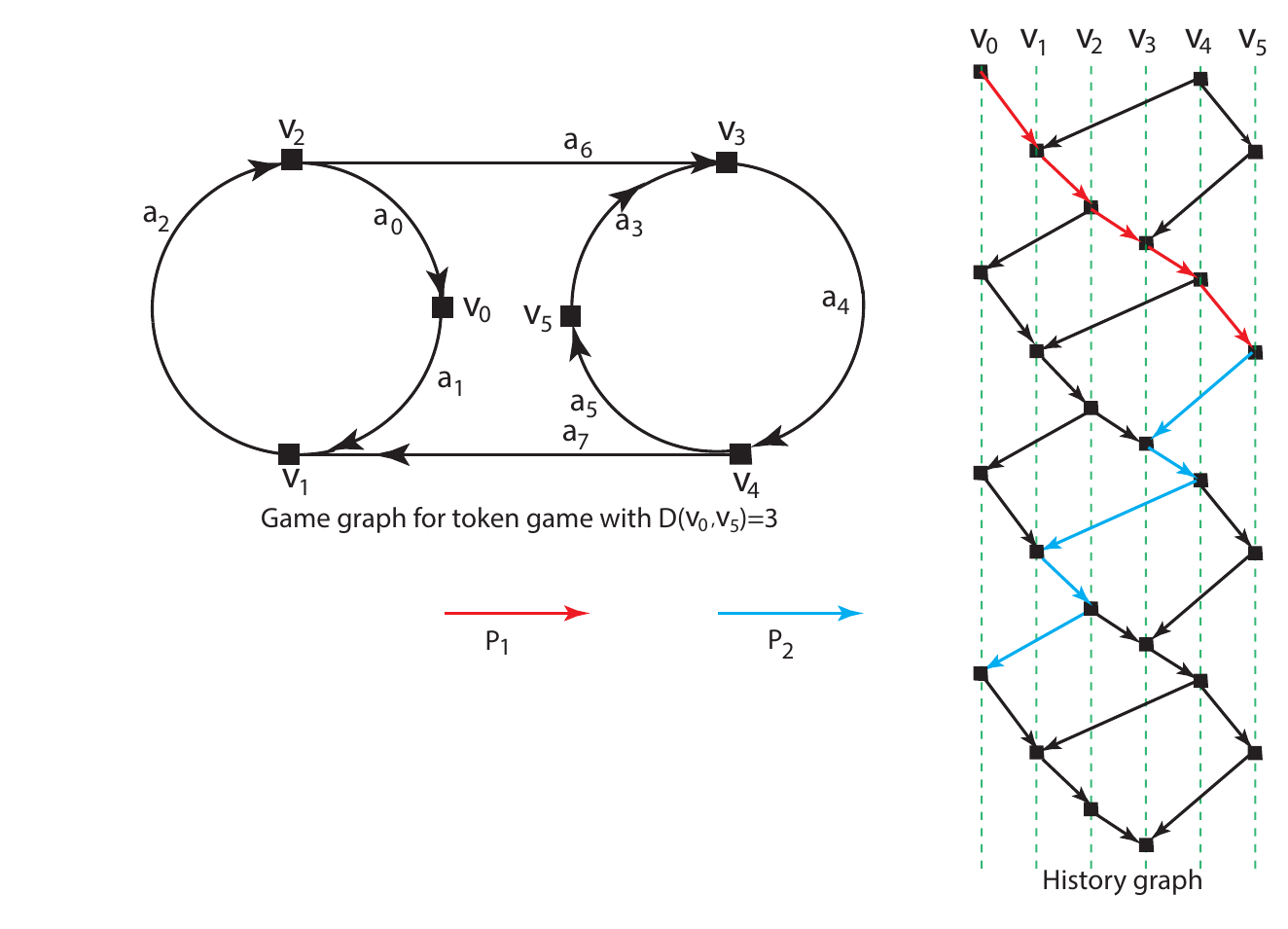}
 \caption{Game graph and history graph of a play of a token game, illustrating logical distance between game-graph nodes $V_0$ and $V_5$.\label{fig:logDist} }
\end{figure}

Note that the paths $P_1$ and $P_2$ on the game graph can overlap, as is the case for those illustrated in \cref{fig:logDist}.

\subsection{Cycle firing counts}\label{sec:4.4}
Proofs in \autocite{71MG} make extensive use of the ``token count'' as the number of tokens in a marking that lie on a cycle of the game graph during a play of a live and safe token game.  By means of partitioned history graphs the same useful number can be obtained without invoking markings.
We proceed in two steps. 
For any cycle  in the game graph of a live and safe  token game, suppose a node of that cycle fires, and, as in the discussion of logical distance, propagates a label around the cycle until it returns to the originating node.  The player of the originating node counts how many times that node fires until the label returns. The count includes the firing that started the label, but not a firing that can occur after the label re-appears on a token on the arrow of the cycle that points into the originating node.  We call this the  {\it cycle firing count}.  The second step is to read the cycle-firing count from the history graph of the play of the token game.  Let $P$ be a spiral path in the partitioned history graph that starts from a firing of a node and proceeds in order through exactly one firing event of each other node on the cycle, returning to a later firing event of the starting node.  Then the cycle firing count for that cycle is the increment of the indices of the events of the originating node from the starting event to the return event.  We have just shown:
\begin{prop}
The number in the usual theory of token games on marked graphs \autocite{71MG}  called the token count of a cycle is the pitch of a spiral in the partitioned history net that projects onto that cycle in the game graph.
\end{prop}

\begin{figure}[H]\hspace*{.8 in}
  %\vspace*{-.3 in} 
 \includegraphics[height=3.5 in]{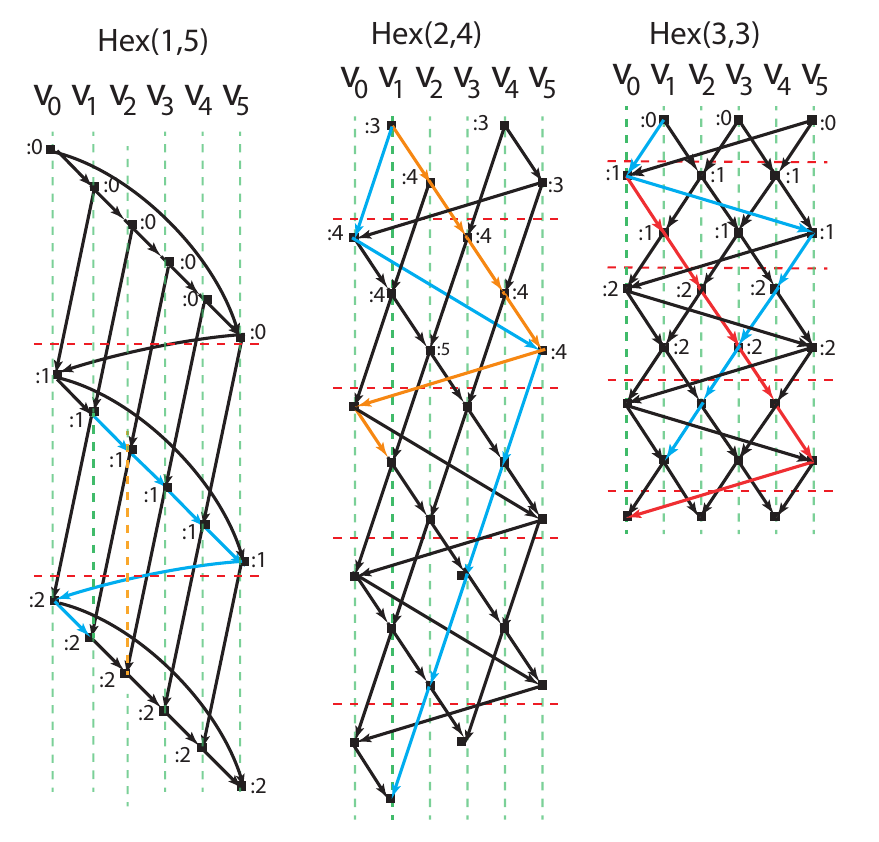}
 \caption{Spiral paths in which each vertex has 1 firing event.\label{fig:paths} }
\end{figure}

For the Hex token games, spirals for cycle-firing counts are illustrated in color in \cref{fig:paths}. The cycle-firing counts can be seen from the number of transverse sectors between the initial events and the return events. Cycles through all six nodes of the game graph of \cref{fig:game} are shown for the three token games that can be played on that graph.
Observe that for Hex(2,4) the cycle (123450) has cycle firing count 2, while going in the other direction, the cycle (1054321) has cycle firing count 4.  For Hex(1,5), the cycle count shown for (123450) is 1; the reverse cycle count, not shown, is 5.  For Hex(3,4), the two cycle counts are both 3.  It is characteristic of game graphs that allow more than one game to have cycles in opposing directions through the same nodes.

%might include paths

\section{History graphs to show changes in live and safe token games}\label{sec:5}
\begin{figure}[H]%\hspace*{.8 in}
  %\vspace*{-.3 in} 
 \includegraphics[height=4.7 in]{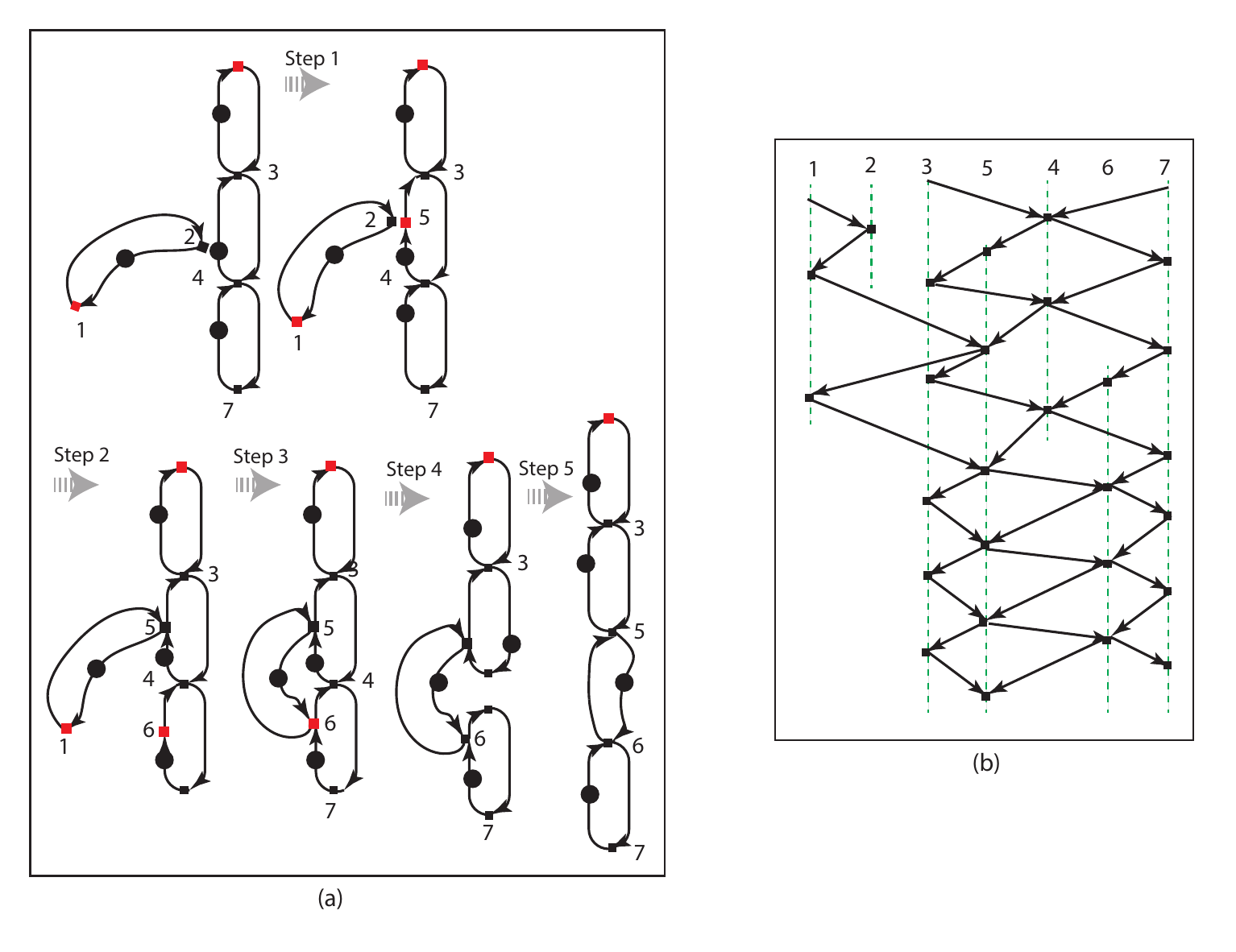}
 \caption{(a) Series of marked graphs undergoing changes while token game continues; (b) single history graph of events of the changing token game\label{fig:histChange} }
\end{figure}

Networks are often in flux, undergoing planned as well as unpredicted changes.  We wondered how to represent the insertion of a new link into a chain of communications over a network without interrupting the chain?  An answer was inspired by a study in biology that showed a series of snapshots as an amoeba inserted itself into a chain of amoeba forming a filament of a cellular slime mold.  In \autocite{25bmb} we translated the series of snapshots into a series of live and safe marked graphs.  Each next graph was related to the previous graph by a change such that the play of the token game could proceed without interruption \autocite{25bmb}.  By means of the history graph illustrated in \cref{fig:histChange}, the node-firing events of a play that proceeds through such changes, predictable or not, can be portrayed not by a series of marked graphs, with their markings,  but by a single history graph without invoking markings.

In \cref{fig:histChange}, the numbers identify nodes, some of which are created
and others of which are terminated during the changes that take place.  In
\cref{fig:histChange}(b) the creation and termination of nodes is visible along
the vertical green dashed lines. The earliest node firings are for a 2-cycle
(1,2) of node firings representing an amoeba disjoint from the filament, which is
characterize by the marked graph prior to step 1.  This marked graph is
characterized here by the nodes 3, 4, and 7.  After the last change, step 5, the
lone amoeba has inserted itself to from the graph to form a chain on the game
graph characterized by the nodes 3, 5, 6, and 7.  Details of the steps from game
graph to game graph that preserve liveness and safety can be found in
\autocite{25bmb}.

\begin{figure}[H]\hspace*{.8 in}
  %\vspace*{-.3 in}
 \includegraphics[height=4 in]{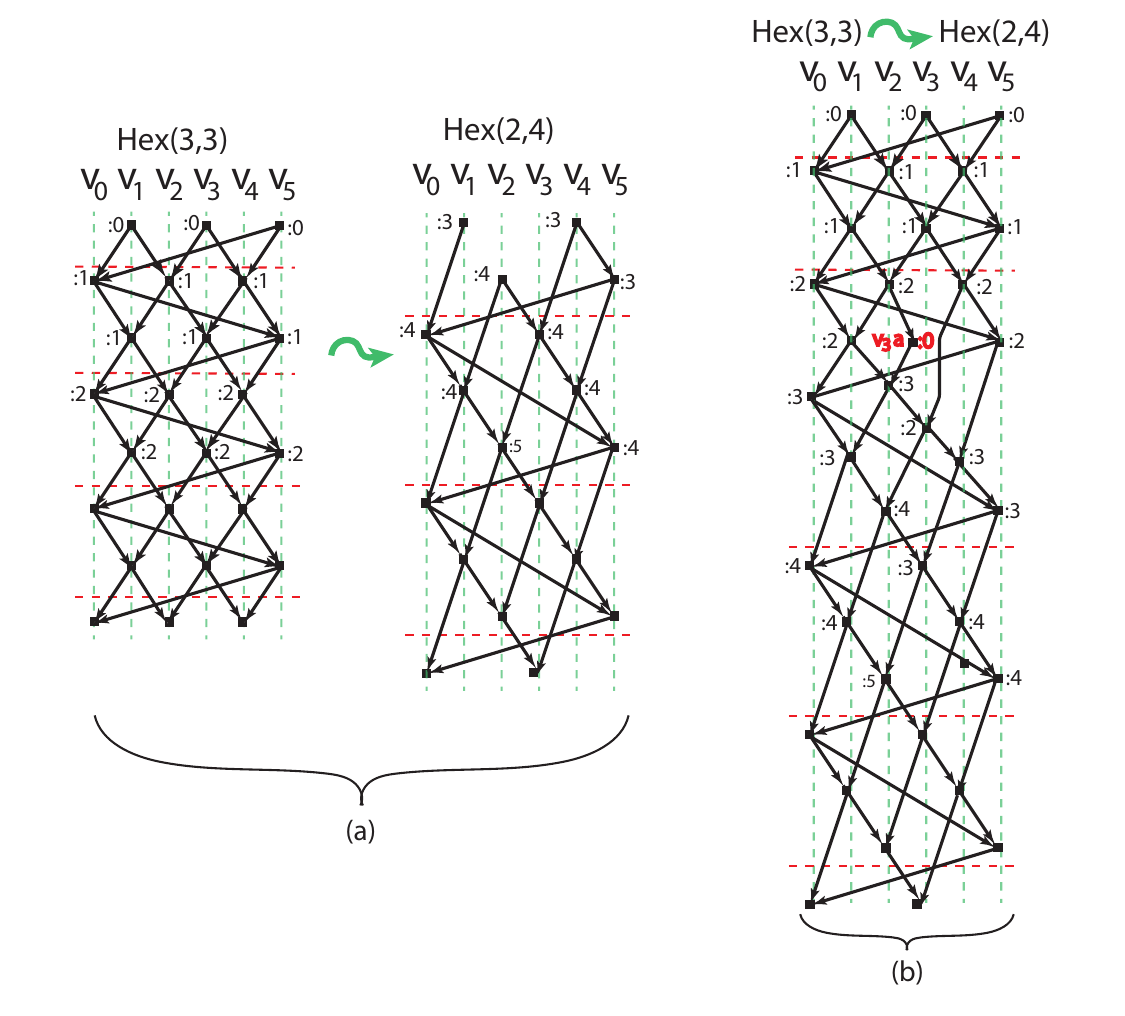}
 \caption{(a) histories for Hex(3,3) and Hex(2,4); (b) history showing a change in the game from Hex(3,3) to Hex(2,4) in the midst of a play. \label{fig:gameChange} }
\end{figure}

Now we show another type of change in which token game changes.  Here the game graph undergoes a change but goes back to its starting form, in such a way that the token game is changed. This is possible because the game graph admits distinct live and safe token games, as evidence by differences in cycle firing counts to the two token games.  Both token games are on the ``Hex'' graph shown in \cref{fig:game}. The two token games, both live and safe, are Hex(3,3) and Hex(2,4), named
 for the cycle firing counts on the outer and inner circuits of Hex game graph.
 For reference, \cref{fig:gameChange}(a) shows history graphs for the Hex(3,3) and Hex(2,4). \Cref{fig:gameChange}(b) is the history graph showing a change from game to the other in the midst of a play.

Here are the details. When $v_3\mc 2$ became fireable, instead of firing it was separated into two nodes $v_3a$ and $v_3b$, each with one arrow pointing in and 1 arrow pointing out. Then  $v_3a$ fires and is a predecessor for  $v2\mc3$ while $v_3b$ never fires (and so has  no firing event in the history graph).  The result of this splitting of the node and the firing of one part is to change the play from the Hex(3,3) game to the Hex(2,4) game.  After the firing $v_3a\mc0$ the two parts of the $v_3$ node are rejoined. Recall  that the indices of any node indicate only the total order of firings of that node, so that any integer added to all the indices for a node has no significance.

The topic of changes in token games that preserve liveness and safety has been explored in a preliminary way (SRL,25bmb,murata, koh), but many questions remain to be formulated, let alone answered.

Because of the change shown in \cref{fig:gameChange}, there is an absence of transverse sector where the change is shown in the partitioned history graph.

\section{Results and discussion}
Sometimes we the authors think in terms of states as global snapshots; they are convenient and carry the habit of tradition.  However, in this report we showed the conceptual deficiency of global snapshots, and we pointed to an alternative based on assemblages of local snapshots.  We showed how local records can be assembled to record all that can reported about distributed networks, and how whatever beyond that can be hoped for from global snapshots is illusory.  Thus, one does not {\it have} to think in terms of global states.

The results here rest on a background  characterized by the mottoes ``no final answers'' and ``no universal clock''.  The two mottoes are linked. Besides the ``no final answers'' implied by quantum theory \autocite{05aop}, there is an older result in physics that speaks against ultimate explanations by asserting ``no universal clock.'' First some history of thought.  It is widely recognized that the Enlightenment was influenced by Newton's innovations in physics, and, in particular, his mathematical expression of absolute time.  Newton's physics with its absolute time led to the resolution of many (but not all) mysteries, and greatly encouraged the tackling of new problems. Absolute time, however, fed into what one might call ``over-enthusiasm'', in the positing an intelligence that can survey the whole world from on high and at once.  Two centuries later, Einstein's relativity offered a correction (which we believe is not yet sufficiently appreciated).
In contrast to  Newton's concept of time and space, special and general relativity assert  a physical limit on the propagation of messages. From a modern point of view, Newtonian physics tells ``how it could be if  the speed of message propagation were infinite.''  Extensive experience with communications, for example in the Global Positioning System (that brings the time of day to our cell phones) confirms that messages do not arrive instantaneously.  For example,  a clock ticks more than half an hour for a message to go to Mars and back, and we can't know what's happening on mars more recently than a quarter hour ago.  History graphs based on local records take account of the experimental fact that no intelligence can know everything at once.

\section*{Acknowledgment}
We thank Judith Clapp for incisive help in organizing our writing.  We thank Susan Wood for helpful criticisms and suggestions.  
\appendix
\section{Appendix: Definitions}\label{sec:A}
A game graph is directed multi-graph, with loops allowed, consisting of:
\begin{enumerate}
\item A set $V$ of elements that we call ``nodes.''
\item  A set $X$ of elements that we call ``arrows.''
\item  A function $s$ whose domain is the set $X$ of arrows and whose range is contained in $V$.
  \item A function $t$ whose domain is $X$ and whose range is contained in $V$.
\end{enumerate}
For any arrow of $X$, the function $s$ assigns a node that we call the ``source,'' and the function $t$ assigns a node that we call the ``target'' of the arrow.  An arrow is {\it incident} with both its source and its target.   In addition,
\begin{enumerate}
\item[5.] The set $V$ is finite and not empty.
\item[6.]  The set $X$ is finite.
\end{enumerate}
The allowance of loops (arrows for which the source and the target are the same node) is motivated by mappings, as discussed in \autocite{SRL}.

A {\it history graph} is an acyclic multi-graph (so without loops) consisting of:
\begin{enumerate}
\item A set $V$ of elements that we call ``events.''
\item  A set $X$ of elements that we call ``edges.''
\item  A function $s$ whose domain is the set $X$ of edges and whose range is contained in $V$.
  \item A function $t$ whose domain is the set $X$ of edges and whose range is contained in $V$.
\end{enumerate}
For any edge of $X$, the function $s$ assigns an event that we call the ``source,'' and the function $t$ assigns an event that we call the ``target'' of the edge.  An edge is {\it incident} with both its source and its target.   In addition,
\begin{enumerate}
\item[5.] The set $V$ is not empty.
\item[6.]  The set $X$ is locally finite, by which we mean that for any two nodes connected by a path, the path is of finite length.   (In \autocite{88BestFer}, `locally finite' is termed `weakly discrete'.) By ``path'' we mean a directed path.
\end{enumerate}
An arrow (for a game graph) or an edge (for a history graph) {\it connects} its source to its target. 

Definitions pertaining to graphs vary among authors.  Ours are the following:\\
A {\it  path} in a game graph is a sequence of arrows,  joined head to tail at nodes, with no repeated arrows.\\  Similarly a path in a history graph is a sequence of edges, etc. By path we mean what often is called a {\it directed path}.
A {\it simple} path has no repeated nodes or edges, respectively.\\
A {\it circuit} is a closed path (which can be a {\it loop})\\ 
  A {\it cycle} is  a closed simple path. (No repeated nodes or edges.)\\
    A game graph is {\it strongly connected}
    if it contains a (directed!) path from every node to every other node. \\

   Although we make no use of them in this report, for reference we state conventional definitions for marked graphs state in terms of markings as global snapshots. 
    \begin{defn}
An {\it unlabeled marking} of a graph is an assignment of a (non-negative)
  whole number, called the {\it token number}, to each arrow. In this report,
  that number is either 0 or 1.  A {\it labeled marking} accompanies an
  assignment of 1 (representing token presence) with a label consisting of a string of 0's
  and 1's (sometimes just a single 0 or 1).
\end{defn}
\noindent A node is {\it fireable} in a marking if the token number on each in-arrow of the node \vspace*{3pt}
is at least 1.\\  
We are interested in two types of marked graph: those with tokens that can be only present or absent from an arrow, and those that, in addition, carry labels.  
For an unlabeled marked graph, a {\it firing} of a node is a relation between two markings: markings $M$ and $M'$ are related by the firing of node $v$ if the token number of $M'$ for each in-arrow of node $v$ is 1 lower than the corresponding token number for $M$, and if the token number  of $M'$ for each out-arrow of node $v$ is  1 higher than the corresponding token number of \vspace*{3pt}$M$.\\
For a labeled marked graph some nodes are given rules that augment firings of nodes by specifying labels on tokens on out arrows as functions of labels on tokens of \vspace*{4pt}
in-arrows.

\printbibliography
\end{document}